\documentclass[12pt,preprint]{aastex}
\begin{document}
\title{Discovery of the Spin Frequency of 4U 0614+09 with {\it Swift}/BAT}
\author{Tod E. Strohmayer} \affil{Astrophysics Science Division,
NASA/GSFC, Greenbelt, MD 20771; stroh@milkyway.gsfc.nasa.gov}
\author{Craig B. Markwardt} \affil{UMD/CRESST/GSFC, Greenbelt, MD
20771; craigm@milkyway.gsfc.nasa.gov} \and \author{Erik Kuulkers}
\affil{ISOC, ESA/ESAC, Urb. Villafranca del Castillo, PO Box 50727,
28080 Madrid, Spain; Erik.Kuulkers@esa.int}
\begin{abstract}

We report the discovery of burst oscillations at 414.7 Hz during a
thermonuclear X-ray burst from the low mass X-ray binary (LMXB) 4U
0614+091 with the Burst Alert Telescope (BAT) onboard {\it Swift}. 
In a search of the BAT archive, we found two burst triggers consistent
with the position of 4U 0614+091. 
We searched both bursts for high frequency timing signatures, and
found a significant detection at 414.7 Hz during a 5 s interval in the
cooling tail of the brighter burst. This result establishes the spin
frequency of the neutron star in 4U 0614+091 as $\approx 415$ Hz.  The
oscillation had an average amplitude (rms) of $14 \%$, These results
are consistent with those known for burst oscillations seen in other
LMXBs. The inferred ratio of the frequency difference between the twin
kHz QPOs, and the spin frequency, $\Delta\nu / \nu_s$, in this source
is strongly inconsistent with either 0.5 or 1, and tends to support
the recent suggestions by Yin et al., and Mendez \& Belloni, that the
kHz QPO frequency difference may not have a strong connection to the
neutron star spin frequency.

\end{abstract}
\keywords{stars: neutron---stars: rotation---stars: oscillations---X-rays:
bursts---X-rays: binaries--X-rays: individual (4U 0614+091)}

\section{Introduction}

The detection of spin modulation pulsations during thermonuclear X-ray
bursts, ``burst oscillations,'' has become an important method for
measuring the spin rates of neutron stars in accreting binary systems
(see Strohmayer \& Bildsten 2006 for a review). To date, almost all of
the {\it a priori} detections of burst oscillations have been made
with the Proportional Counter Array (PCA) onboard the {\it Rossi X-ray
Timing Explorer} (RXTE). Indeed, its combination of large collecting
area, low background, high time resolution, and high telemetry
capacity enabled the first detections of the phenomenon. Pulsations
during a bright burst from SAX J1808.4-3568 were detected with the
Wide Field Camera (WFC) on {\it BeppoSAX} at the $\approx 3\sigma$
level, but in this case the spin frequency was already known (in 't
Zand et al.  2003).

The Burst Alert Telescope (BAT) onboard {\it Swift} has a combination
of large collecting area (5,200 cm$^2$) and high time resolution (100
$\mu$-sec) that should, in principle, make it sensitive to burst
oscillations from particularly bright X-ray bursts, although the BAT
does suffer a significant loss of sensitivity to such oscillations due
to its much larger background compared to the PCA. The BAT has a
sophisticated triggering system that identifies Gamma-ray bursts
(GRBS), and other fast transients, including thermonuclear bursts from
accreting neutron stars. When a trigger occurs, an estimate of the
source's position is derived onboard the {\it Swift} spacecraft, where
it is checked against an onboard catalog of known objects. If the
position is consistent with a source in the catalog, then the trigger
is deemed to not be a GRB, and {\it Swift} will not initiate a direct
slew to the position. However, high time resolution data is still
accumulated around the trigger time. These data can then be searched
for burst oscillations.

Historically, some of the brightest X-ray bursts observed, with fluxes
of order $2 \times 10^{-7}$ ergs cm$^2$ s$^{-1}$, have been associated
with the low mass X-ray binary (LMXB) 4U 0614+091. Based on the
detection of a bright burst with {\it Watch}, and Eddington limit
arguments, Brandt et al. (1992) argued that the distance to 4U
0614+091 is likely $< 3$ kpc. Several lines of evidence point to the
conclusion that 4U 0614+091 is an ultra-compact X-ray binary system.
Juett, Psaltis \& Chakrabarty (2001) inferred an enhanced neon to
oxygen ratio for the system, and based on a comparison of the ratio
with known ultra-compact systems, argued that 4U 0614+091 is a similar
system. Sensitive optical spectroscopy by Nelemans et al. (2005) found
carbon and oxygen emission lines, but no evidence for helium or
hydrogen, further strengthening the ultra-compact classification.  The
optical counterpart to 4U 0614+091, V1055 Ori, is also intrinsically
faint, consistent with a compact orbit (Nelemans et al. 2005; van
Paradijs \& McClintock 1994).  The faintness of its persistent X-ray
emission and nearby distance suggest a low accretion rate, again
consistent with an orbital period $< 1$ hr (Deloye \& Bildsten
2003). Finally, we note that in 't Zand et al. (2007) cite a report by
O'Brien (2005) of a 50 min. optical modulation, that could represent
the orbital period.

Although 4U 0614+091 has been extensively studied with RXTE, it has
not yet observed any X-ray bursts from the source.  Ford et al. (1997)
reported the discovery of kHz quasiperiodic oscillations (QPOs) from
the source using RXTE data. They found a maximum QPO frequency of
$\approx 1,145$ Hz, and a mean frequency separation of $323 \pm 4$ Hz
between the upper and lower kHz oscillations (see van der Klis 2006,
for a review of kHz QPOs). Based on this frequency separation they
suggested a spin frequency of $\approx 323$ Hz.  Recent studies,
however, have suggested that the frequency separation between the kHz
QPOs may be more or less independent of the neutron star spin
frequency (Yin et al. 2007; Mendez \& Belloni 2007).

In this Letter we report the discovery of 415 Hz burst oscillations in
a burst from 4U 0614+091 observed with the BAT instrument onboard
{\it Swift}.  In \S 2 we describe our timing study of bursts from 4U
0614+091 with BAT and summarize their basic properties. In \S 3 we
discuss our results in the context of recent efforts to understand the
relationship (if any) between the kHz QPOs and the spin frequencies of
neutron stars in LMXBs.

\section{Observations and Data Analysis}

We searched the BAT archives for triggered events consistent with the
position of 4U 0614+091, and found two; triggers 234849, and
273106. These events ocurred on October 21, 2006 and March 30, 2007,
respectively. For imaging the BAT has a point spread function of 22
arcmin, but centroiding for source localization is typically accurate
to 1 - 6 arcmin, based on the detection significance. The BAT-derived
positions for each trigger were offset by 1.5 and 1.0 arcmin,
respectively, from the well known position of 4U 0614+091, confirming
it as the source of the bursts.  The BAT data consist of X-ray event
times with 100 $\mu$-sec resolution as well as energy measurements for
43 seconds around each burst.  BAT's nominal energy band is 15 - 150
keV, and the bursts are not detected above about 30 keV, which is
consistent with a thermonuclear origin. Figure 1 shows mask-weighted
(background subtracted) lightcurves of each burst in the 12 - 25 keV
energy band.  The 2006 October 21 burst was observed at $\approx
20.6^{\circ}$ from the BAT pointing axis, so the source countrate was
substantially higher for this event than the 2007 March 30 burst,
which was at $51.3^{\circ}$. Indeed, the peak 12 - 20 keV raw
countrates were 4,100 and 2,620 s$^{-1}$ for the 2006 October and 2007
March bursts, respectively. Time resolved spectroscopy of these bursts
indicates that they both had peak fluxes of $\approx 2 \times 10^{-7}$
ergs cm$^2$ s$^{-1}$ (Kuulkers et al. 2007, in preparation). These
values are comparable to peak fluxes reported from earlier burst
observations (Brandt et al. 1992; Swank et al. 1978). There is no
strong indication for photospheric radius expansion in either burst.

The signal to noise ratio, $n_{\sigma}$, in a timing signature (such
as a QPO) is proportional to $S^2 / (S + B)$, where $S$ and $B$ are
the source and background countrates, respectively. Thus, to search
most sensitively for an oscillation signal one should attempt to
minimize the background and maximize the source countrate.  With the
present BAT data we do this by restricting our timing study to events
in the energy band from 13 - 20 keV. We selected this range because it
effectively maximized the ratio of burst countrate to total
countrate. That is, most of the events above 20 keV are not associated
with the thermal burst emission, and simply add background, while the
BAT does not have much effective area below about 15 keV.  To search
for burst oscillations we computed power spectra for 10 second
intervals, and we stepped the interval by 2 seconds through each
burst. We searched in the frequency band from 1 - 2048 Hz, at
successive resolutions of 0.2, 0.4, and 0.8 Hz. We began our study
with the 2006 October burst since this burst had the higher observed
countrate of the two events.

Our study revealed a significant signal at 415 Hz in the cooling tail
of the 2006 October burst.  Figure 2 shows a power spectrum computed
from a 10 s interval beginning 24.5 s from the beginning of the event
data.  The frequency resolution is 0.2 Hz, and a strong peak in the
power spectrum is present at 414.75 Hz.  The single trial probability
to obtain a peak this strong (25.78) in this power spectrum is $1.7
\times 10^{-10}$. We searched 10,240 frequency bins in this power
spectrum (to 2048 Hz at 0.2 Hz resolution), and a total of 14 similar
power spectra through the burst tail. To account for the number of
trials searched we take the total number of power spectral bins in
each time interval, and at each resolution.  This gives $10,240 \times
14 \times (1 + 0.5 + 0.25) = 250,880$ trials, and yields a detection
significance of $4.3 \times 10^{-5}$, which is a 4$\sigma$
detection. We note that this is a conservative estimate of the
significance, since not all the bins at 0.4 and 0.8 Hz resolution are
independent of those in the 0.2 Hz spectrum. The pulsed amplitude
associated with the signal in this power spectrum is $12.3\%$ (rms).
This is the fractional modulation associated with the burst emission
only, under the assumption that the pre-burst level is a reasonable
indicator of the non-burst background.

We next computed a dynamic power spectrum through the burst using the
$Z^2$ statistic (see Strohmayer \& Markwardt 2002, for a discussion of
the method). We used 4 s intervals, and stepped the intervals at 0.25
s. The dynamic spectrum is shown in Figure 3 and indicates that the
oscillation signal is present during an $\approx 5$ s interval
beginning about 22 s after the onset of the burst. The vertical dashed
lines in Figure 3 mark the interval used to compute the power spectrum
shown in Figure 2.  The presence of oscillations after the burst peak,
during the cooling tail, is consistent with other burst oscillation
sources, and further supports the interpretation that we are seeing
burst oscillations in 4U 0614+091.  The oscillation amplitude is also
within the characteristic range seen in other sources.

We next searched the March 2007 burst in a manner similar to that
described above, however, having detected a signal in the October 2006
burst, we narrowed the search in frequency to within a few Hz on
either side of 415 Hz.  We did not detect any significant signal.  We
are not too suprised by this, however, since the peak burst countrate
is almost a factor of 3 lower in the March 2007 event than the October
2006 burst. This greatly reduces the sensitivity to pulsations in this
burst. Quantitatively, $S^2 / (S + B)$ is reduced by a factor of $\sim
5$, which would erase a $4\sigma$ detection, thus a non-detection is
not particularly surprising.

\section{Discussion}

We have found strong evidence for a burst oscillation at 414.75 Hz in
a thermonuclear burst from 4U 0614+091 with the BAT onboard {\it
Swift}. This is the first detection of burst oscillations with BAT and
demonstrates its potential in this regard, particularly for relatively
nearby (read bright) burst sources.  This detection implies a spin
frequency of about 415 Hz for the neutron star in this system.  Since
the discovery of kHz QPOs and burst oscillations, there has been a
suggested link between the neutron star spin frequency, $\nu_s$, (as
indicated by burst oscillation frequencies, or persistent pulsation
frequencies, or both) and the frequency difference of the twin kHz
QPOs, $\Delta\nu / \nu_s$. Initial discoveries suggested that the kHz
QPO frequency separation was consistent with the spin frequency, but
subsequent observations have revealed a more complex picture
(Strohmayer et al. 1996; van der Klis 2006). The tentative consensus
at present is that the frequency separation is related to the spin
frequency (at least in some sources), in the sense that $\Delta\nu /
\nu_s$ can be close to either 1 or 0.5.  While this description fits
reasonably well with several of the observed systems, it is not
universally precise, in the sense that some objects show statistically
significant deviations from 1 or 0.5, and that the frequency
separation is known to vary in some objects. Recently, Mendez \&
Belloni (2007), and Yin et al. (2007) have re-examined the existing
measurements of neutron star spin frequencies and kHz QPO, and suggest
that there may be only a weak connection between the QPO frequency
difference and the spin frequency.  Indeed, Mendez \& Belloni (2007)
hypothesize that the frequency difference may be more or less
constant, essentially independent of the spin frequency.

With the detection of the spin frequency in 4U 0614+091 we now have an
additional source in which to closely evaluate the relationship
between kHz QPOs and spin. Interestingly, the burst oscillation
frequency we have found is very significantly different from the
frequency separation of $323 \pm 4$ Hz of the twin kHz QPOs reported
for the system (Ford et al. 1997).  In this case the QPO difference
frequency was quite accurately determined, and since the spin
frequency is not thought to differ from observed burst oscillation
frequencies by more than a few Hz, this is a strong demonstration that
the kHz QPO difference frequency and spin frequency in 4U 0614+091 do
not follow the ``consensus'' model.  Assuming that the spin frequency
is within 4 Hz of our measured burst oscillation frequency, then we
arrive at $\Delta\nu / \nu_s = 0.78 \pm 0.02$, which is inconsistent
with both 0.5 and 1 at very high confidence.  Interestingly, this
value lies much closer to the two alternative relations suggested by
Yin et al.  (2007, $\Delta\nu = 390 \; {\rm Hz} - 0.2\nu_s$), and
Mendez \& Belloni (2007, $\Delta\nu = 308 \; {\rm Hz}$), although it
stills fall about $2\sigma$ away from these as well.  Since these
relations were derived and or suggested prior to our discovery, the
spin meausurement reported here can be thought of as an independent
check on those models, and it appears they provide a much better
description of the data than the ``consensus'' model, further
supporting the notion that it may be past time to re-think the
relationship between the kHz QPOs and neutron star spin, or to at
least adopt the working hypothesis that the relationship does not hold
for all sources, or at all times.

\section{Summary}

We have found the first evidence for burst oscillations in the LMXB 4U
0614+091 at 414.75 Hz, the first such detection with the BAT
instrument onboard {\it Swift}.  The inferred spin frequency of
$\approx 415$ Hz is fairly typical for such sources, but it is
inconsistent with the kHz QPO frequency separation, or half of that,
measured for the source by Ford et al. (1997).


\pagebreak

\begin{figure}
\begin{center}
\includegraphics[width=6in, height=6in]{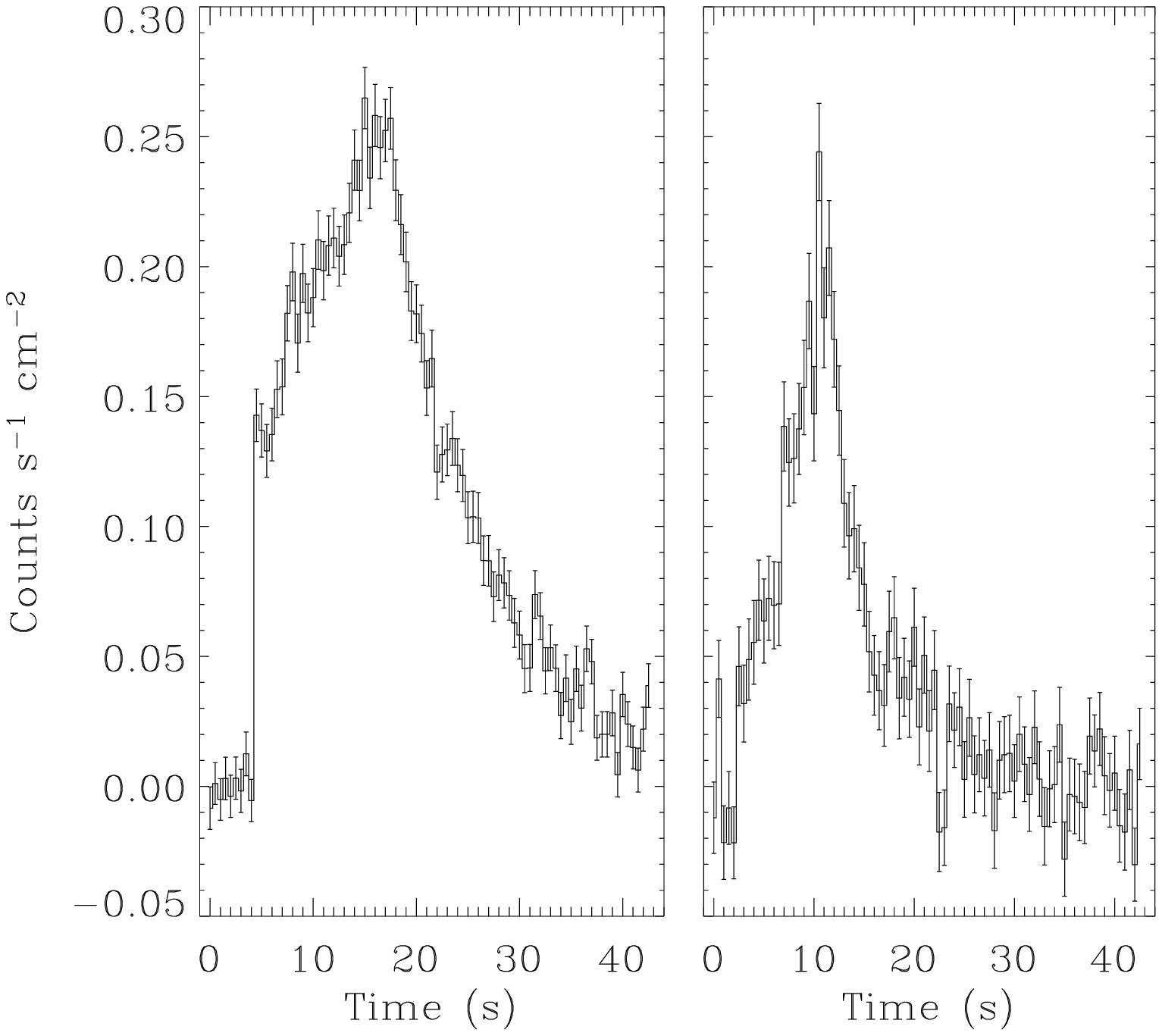}
\end{center}
Figure 1: Mask-weighted light curves in the 12 - 25 keV band of the
two thermonuclear bursts observed by BAT from 4U 0614+091. The left
panel shows the burst from 2006 October (trigger \#234849), and the
right panel that from 2007 March (trigger \#273106). The time bin size
is 0.5 s. Mask-weighted BAT light curves have the background
subtracted, and are corrected approximately to the on-axis effective
count rate.
\end{figure}
\clearpage

\begin{figure}
\begin{center}
\includegraphics[width=6in,height=6in]{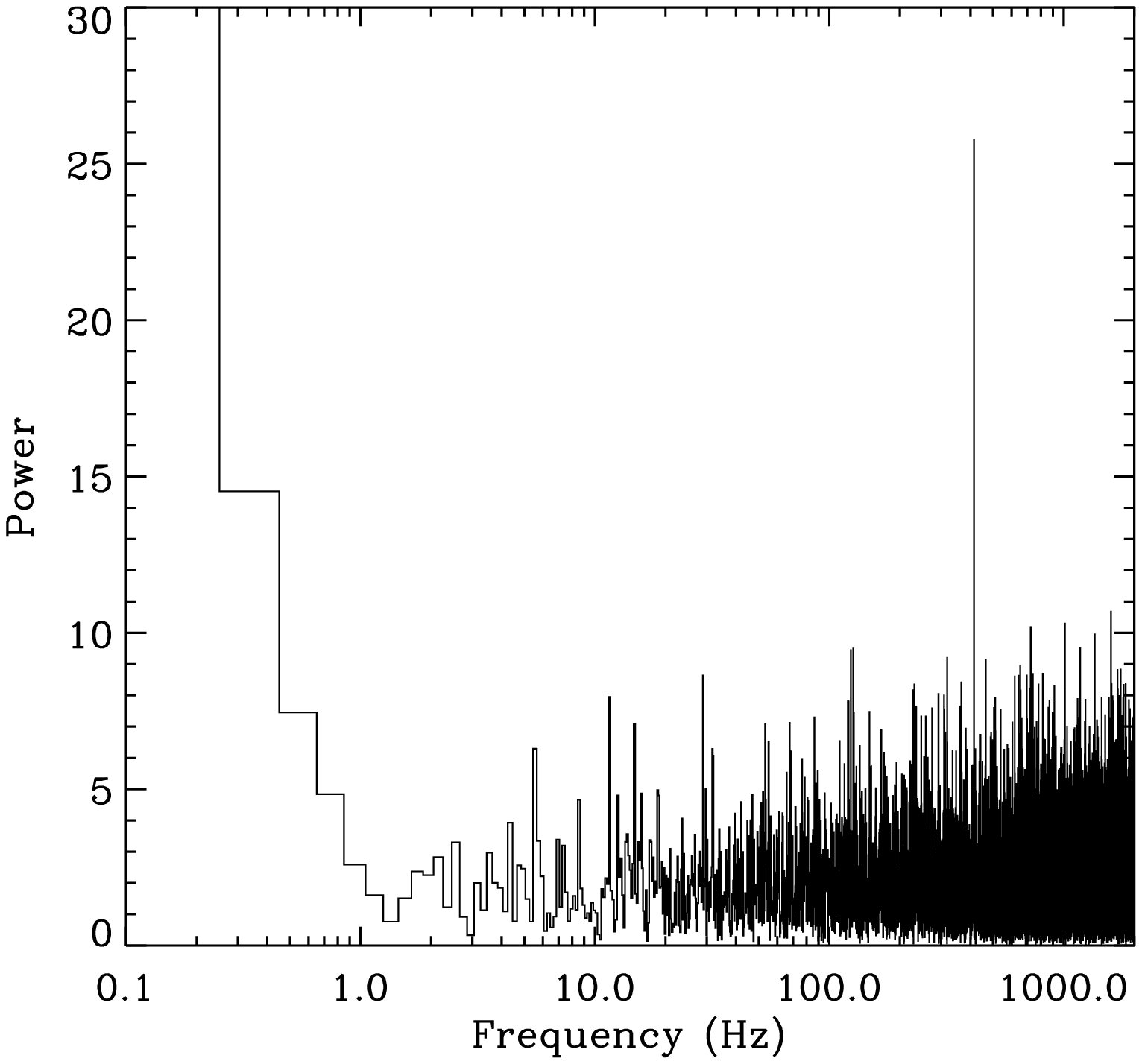}
\end{center}
Figure 2: Power spectrum from the decaying tail of the 2006 October
burst. The spectrum was computed from a 10 s interval beginning 22.5 s
from the start of the data stream (time zero in Figures 2 and 3). For
this spectrum we used events in the 13 - 20 keV band, and the
frequency resolution is 0.2 Hz. The strong peak at $\approx 415$ Hz is
evident.  See the text for a detailed discussion.
\end{figure}
\clearpage

\begin{figure}
\begin{center}
\includegraphics[width=6in, height=6in]{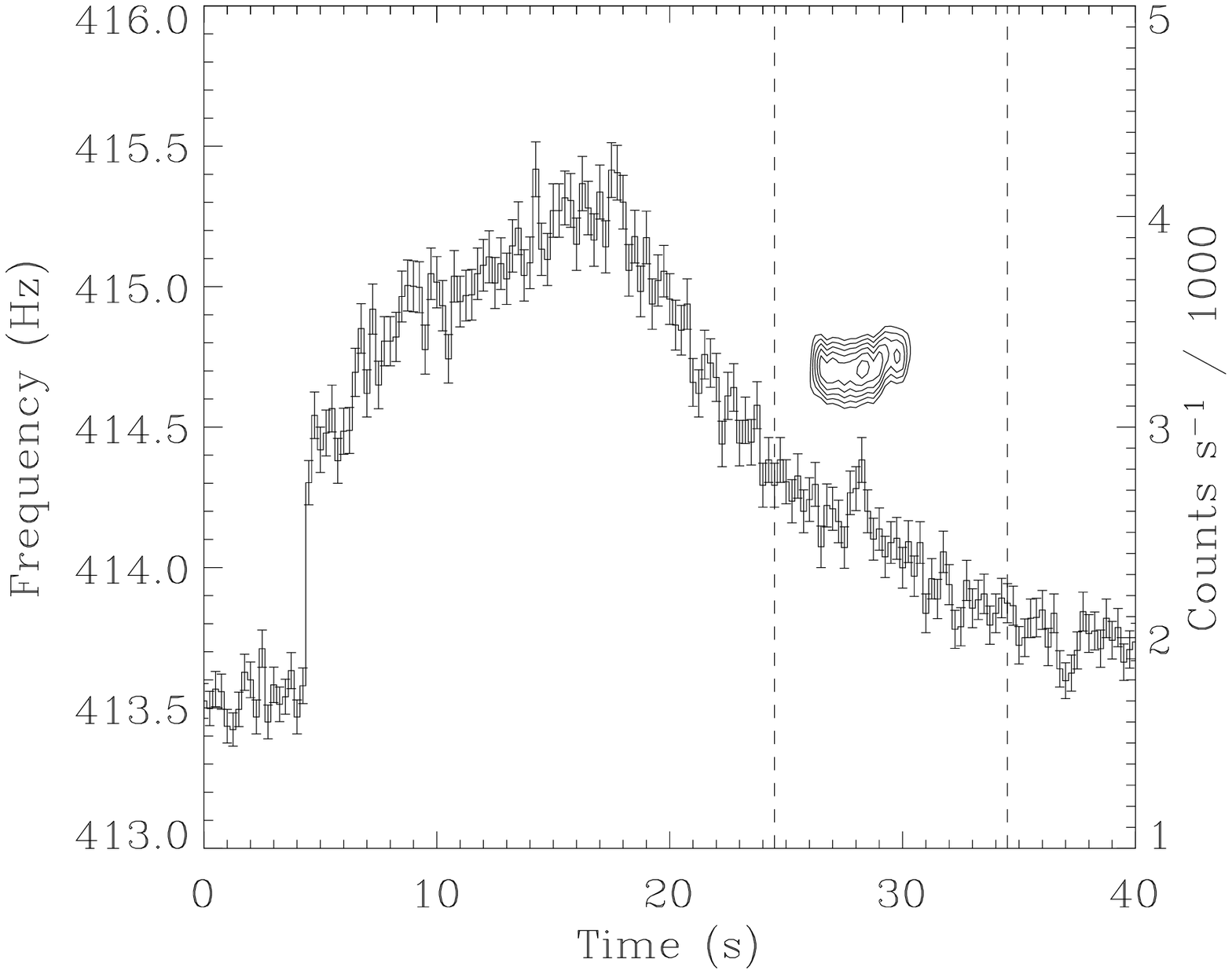}
\end{center}
Figure 3: Dynamic $Z^2$ power spectrum of the burst from 2006
October. We computed $Z_1^2$ power spectra using 4 s intervals, and
stepped the intervals by 0.25 s. We used only events in the 13 - 20
keV range. Six $Z_1^2$ contour levels are plotted, starting at 16 and
spaced in steps of 4. The vertical dashed lines denote the time
interval used to compute the power spectrum displayed in Figure 2. The
burst light curve in the 13 - 20 keV band with 0.25 s time bins is also
superposed. The pulsation signal is evident in the cooling tail of the
burst.
\end{figure}
\clearpage

\end{document}